\PassOptionsToPackage{table, dvipsnames}{xcolor}

\documentclass{vgtc}

\ifpdf \usepackage{graphicx}                \pdfoutput=1\relax                   \pdfcompresslevel=9                  \pdfoptionpdfminorversion=7          

  \DeclareGraphicsExtensions{.pdf,.png,.jpg,.jpeg} \else 
  \usepackage{graphicx}                \DeclareGraphicsExtensions{.eps}     \fi 

\graphicspath{{figures/}{pictures/}{images/}{./}} 

\usepackage{wrapfig}
\usepackage{microtype}                 \PassOptionsToPackage{warn}{textcomp}  \usepackage{textcomp}                  \usepackage{mathptmx}                  \usepackage{times}                              \usepackage{cite}                      \usepackage{tabu}                      \usepackage{booktabs}                  

\usepackage{graphicx}

\usepackage{url}

\usepackage{breakurl}
\usepackage[breaklinks]{hyperref}

\usepackage{hyperref}
\hypersetup{
    colorlinks=true,
    linkcolor=blue,
    filecolor=magenta,      
    urlcolor=cyan,
    pdftitle={Overleaf Example},
    pdfpagemode=FullScreen,
    }

\usepackage{soul}

\onlineid{0}

\vgtccategory{Research}

\vgtcinsertpkg

\title{Only YOU Can Make IEEE VIS Environmentally Sustainable}

 \makeatletter
\def\@fnsymbol#1{}
    \makeatother

\newcommand{\emojiElsie}{{\includegraphics[width=0.1in]{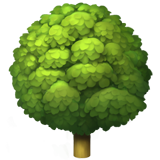}}}
\newcommand{\emojiAndrew}{{\includegraphics[width=0.1in]{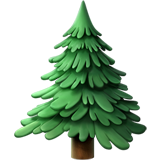}}}

\author{Elsie Lee-Robbins~\emojiElsie{}\thanks{\emojiElsie{}~e-mail: elsielee@umich.edu}\\ \scriptsize University of Michigan \and Andrew McNutt~\emojiAndrew{}\thanks{\emojiAndrew{}~e-mail: mcnutt@uchicago.edu}\\ \parbox{1.4in}{\scriptsize \centering University of Chicago}}

\newcommand{\etal}
{et al.}

\newcommand{\ie}{{i.e.,}}
\newcommand{\eg}{{e.g.,}}

\newcommand{\figref}[1]{\hyperref[#1]{Fig.~\ref*{#1}}}
\newcommand{\secref}[1]{\hyperref[#1]{Sec.~\ref*{#1}}}

\newcommand{\parahead}[1]
{\paraheadd{#1}.
}

\newcommand{\paraheadd}[1]
{\vspace{0.07in}\noindent \textbf{\textit{#1}}}

\def\subsubsec#1
{\subsubsection{#1}}

\newcommand{\hlc}[2][yellow]{{\colorlet{foo}{#1}\sethlcolor{foo}\hl{#2}}}
\newcommand\qt[1]{\hlc[SeaGreen!15]{``#1''}}

\usepackage[normalem]{ulem}
\definecolor{linkColor}{HTML}{257E98}
\setuldepth{Berlin}
\newcommand\asLink[2]{\textcolor{linkColor}{\href{#1}{\ul{#2}}}}

\newcommand{\osf}{\asLink{https://osf.io/u2eps/}{osf.io/u2eps/}} 
\abstract{

The IEEE VIS Conference (or VIS) hosts more than 1000 people annually. It brings together visualization researchers and practitioners from across the world to share new research and knowledge.
Behind the scenes, a team of volunteers puts together the entire conference and makes sure it runs smoothly. Organizing involves logistics of the conference, ensuring that the attendees have an enjoyable time, allocating rooms to multiple concurrent tracks, and keeping the conference within budget.
    In recent years, the COVID-19 pandemic has abruptly disrupted plans, forcing organizers to switch to virtual, hybrid, and satellite formats.
    These alternatives offer many benefits:  fewer costs (\eg{} travel, venue, institutional), greater accessibility (who can physically travel, who can get visas, who can get child care), and a lower carbon footprint (as people do not need to fly to attend).
As many conferences begin to revert to the pre-pandemic status quo of primarily in-person conferences, we suggest that it is an opportune moment to reflect on the benefits and drawbacks of lower-carbon conference formats.   
To learn more about the logistics of conference organizing, we talked to 6 senior executive-level VIS organizers. We review some of the many considerations that go into planning, particularly with regard to how they influence decisions about alternative formats.  
    We aim to start a discussion about the sustainability of VIS---including sustainability for finance, volunteers, and, central to this work, the environment---for the next three years and the next three hundred years. 
}


\nocopyrightspace

\begin{document}

\firstsection{Introduction}

\maketitle

What is the environmental impact of academic travel? What are  ways to reduce a conference's carbon footprint? Do we even care to make these changes?

\begin{figure}[t]
  \includegraphics[width=\linewidth]{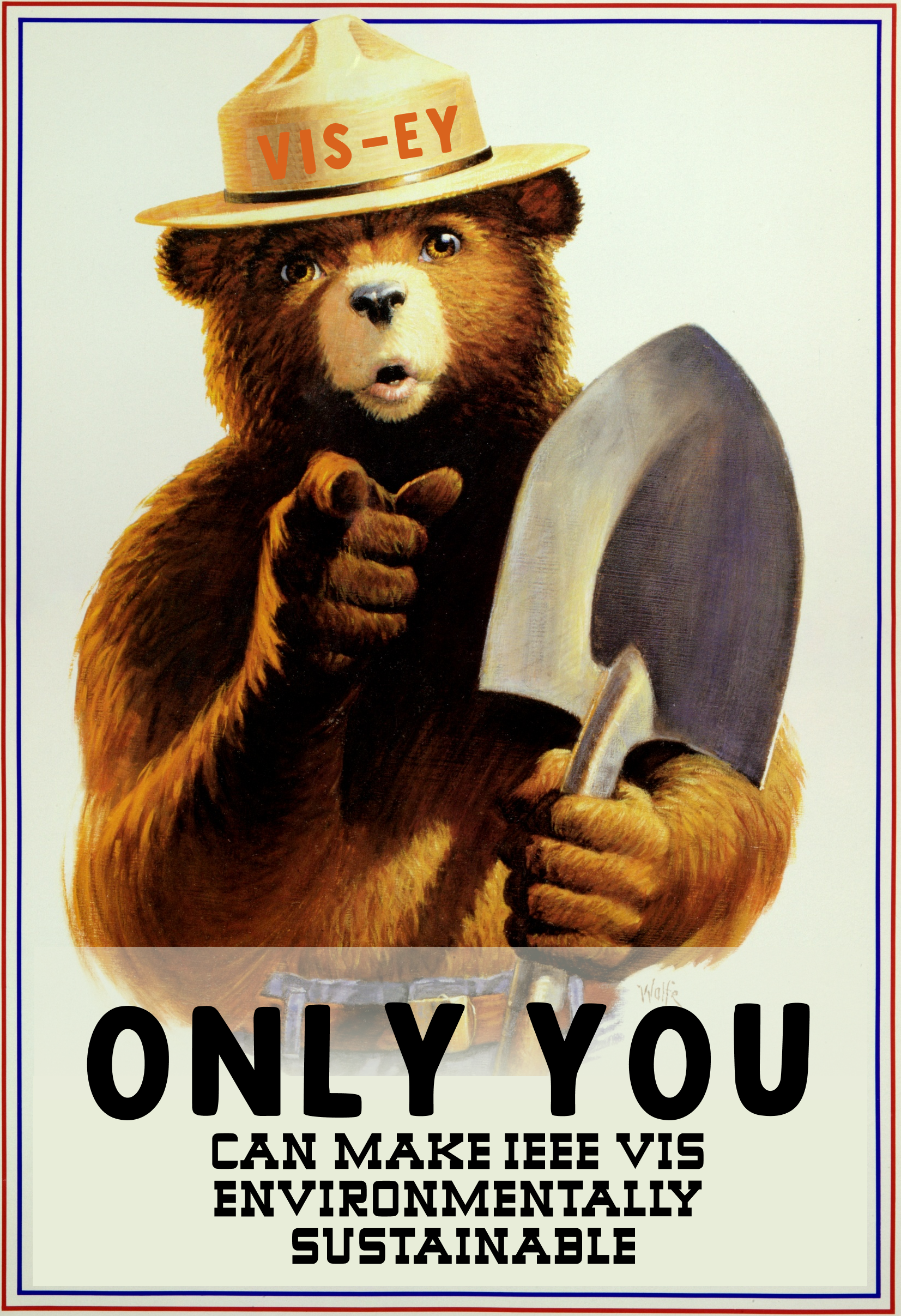}
  \caption{Smokey (the) bear is a mascot of the U.S. Forest Service that advocates personal responsibility in preventing wild fires. Here he has graciously changed his message to urge you to take ownership over making IEEE VIS environmentally sustainable.  }
  \vspace{-1em}
  \label{fig:smokey}
\end{figure}

Climate change is an important and urgent issue.
As we write this paper, we are experiencing the hottest day in recorded history (July 4, 2023)~\cite{washingtonpost2023}---a title that day will not hold for long.
The Intergovernmental Panel on Climate Change (IPCC)
stated that:
``{Climate change is a threat to human well-being and planetary health. There is a rapidly closing window of opportunity to secure a liveable and sustainable future for all}''~\cite{panelOnClimateChange2023}.
Action needs to be taken this decade (and as soon as possible) to ensure that life on Earth is at least livable for everyone on it.
Unfortunately, due to the slow and global scale of environmental change, it is easy to dismiss climate change as a distant problem that is someone else's responsibility~\cite{gore2006inconvenient}, or as something individuals are exempt from as it can only be dealt with on a governmental or institutional scale.
Just as U.S. Forest Service mascot Smokey Bear advises that ``\textit{Only YOU can prevent wildfires},'' we believe it is every VIS participant and organizer's responsibility to make VIS environmentally sustainable (\figref{fig:smokey}).

Institutional-scale behaviors are made up of individual actions.
Global air travel makes up $\approx$2.5\% of total global CO$_{2}$ emissions and $\approx$3.5\% of global warming~\cite{Ritchie20avaiation}.
Consider a provocative calculation: a round-trip flight from Seattle to Melbourne (the location of IEEE VIS 2023) is $\approx$2.7 tons of CO$_2$ emissions~\cite{GuardianFlights}.
For scale, that is a bit more than ``{driving from Boston to New Delhi in India and back (assuming you have a car that can drive on water)}''~\cite{mitClimateCarbon}.
An international flight could easily be your largest carbon emission for an entire year: more than your car emissions, the food you eat, or the things you buy.

For better or worse, Computer Science has primarily tied publications to conference attendance in a way that is unlike other research areas. Other disciplines also share research at conferences, but as works-in-progress and not as the final, archival publication. Computer Science conferences are, as Vardi describes it, a ``journal that meets in a hotel''~\cite{vardi2019publish}.
This causes our conferences to have large carbon footprints, as scholars from across the world are asked to fly to a wide variety of (often not nearby) locations as a key part of their work~\cite{berne2022carbon}, sometimes multiple times per year.

However, a large carbon footprint is not unavoidable.
As a result of the pandemic, alternative conference formats (such as remote and hybrid) were briefly widespread, significantly reducing the conference carbon footprint and making those conferences more accessible (as a visa or physical accommodation was not required to attend).
As pre-COVID practices return, it seems likely that such affordances will end, with venues such as UIST, VIS, and CHI electing to offer no or limited hybrid options.

\begin{figure}[t]
  \vspace{-2em}
  \includegraphics[width=\linewidth]{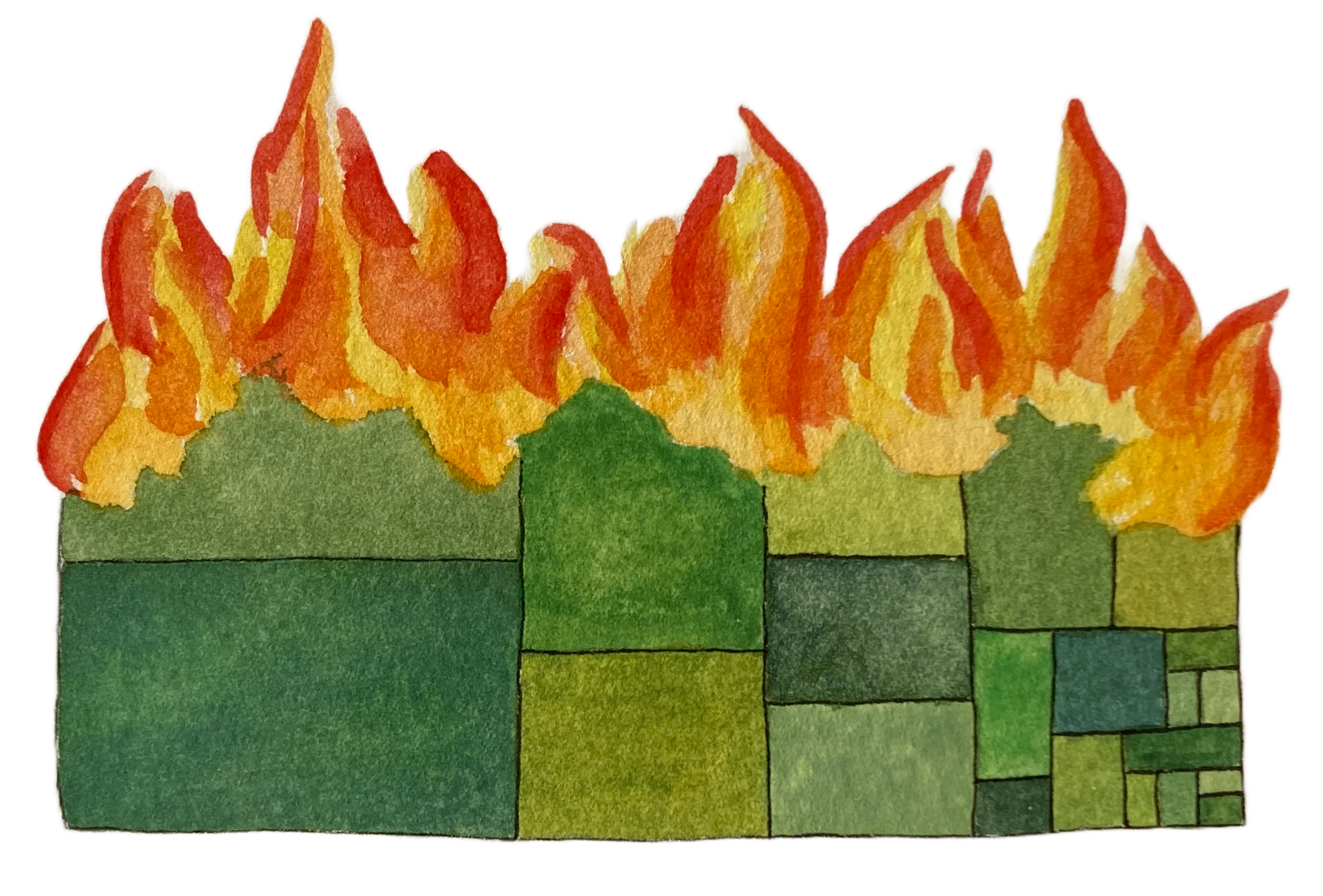}
  \vspace{-4em}
\end{figure}

It is  easy (and enjoyable) to return to the status quo~\cite{mcnutt2021villainy}, as the benefits of in-person experiences are visible and visceral, but we are also offered an opportunity to reflect and consider what is essential about VIS.
Should the conference strive to consider its environmental impacts?
How much should we value alternative conference formats as a way to lower our carbon footprint?
As IEEE VIS is the ``{premier forum for advances in visualization and visual analytics}''~\cite{ieeeVIS}, and hence an institution with the power and impact that comes with spending millions of dollars and involving thousands of people, how do we wish to exert that power?

In this paper, we explore the benefits and drawbacks of alternative conference formats that would lead to a lower carbon footprint.
We center this discussion around the considerations that go into the immense task of organizing a 1200-person event~\cite{besanccon2022dystopia}.
To learn more about how VIS is organized, we conduct an interview study with ($N=6$) executive-level organizers of the IEEE VIS conference.
We uncover a variety of trade-offs and perspectives, and find there is no perfect solution for how we should engage with sustainability.
Our goal in this work is to raise the issue of environmental sustainability in our community.
To support this dialog, we sketch a variety of ways to make VIS  more sustainable, such as via remote, hybrid, and regional conferences.
As ours are just two voices, we do not aim to give a recommendation of which conference format is the best, but rather to start a conversation about the sustainability of IEEE VIS.

\definecolor{Mycolor2}{HTML}{F2994A}

\section{IEEE VIS and Climate Change}

\begin{figure}[t]
  \includegraphics[width=\linewidth]{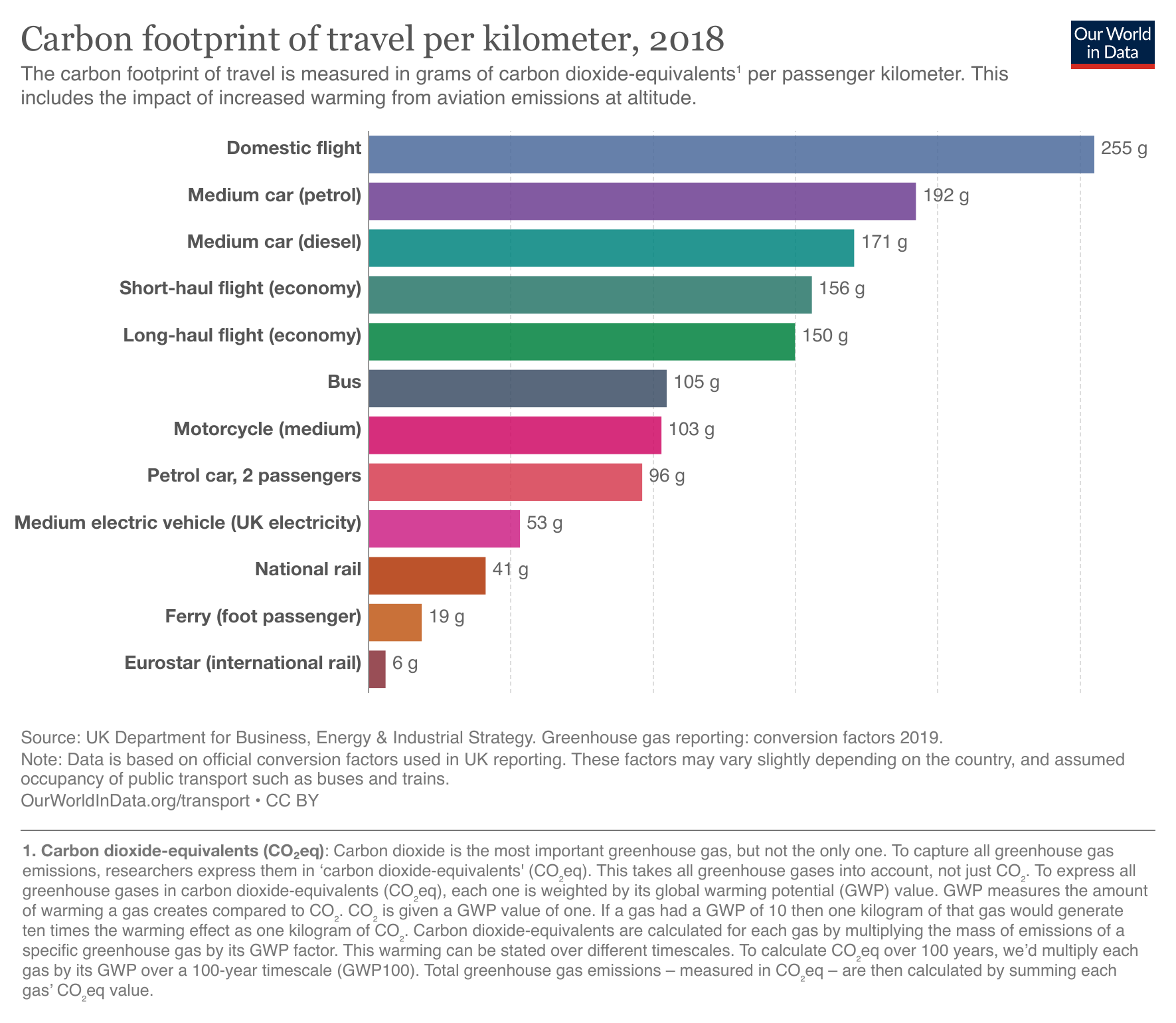}
  \caption{The carbon footprint of different modes of travel. How might consideration of how attendees travel to the conference influence how we organize it? Chart created by Hannah Ritchie from Our World in Data \cite{ritchie2020transport}.}
  \label{fig:transport}
\end{figure}

This paper is far from the first time VIS has interacted with the climate. For instance, the Viz4Climate workshop~\cite{viz4Climate} brought together visualization researchers and climate scientists. Visualizations can be used to educate the public on how to be more sustainable (\eg{} Fig. \ref{fig:transport}, \cite{ritchie2020transport}) and inspire conversations about climate change (\eg{} \cite{HawkinsStripes}). Instead, we turn this question and ask how are we \textit{contributing} to climate change?

Academic travel is, if not essential to, a large part of career development. Publish (and present) or perish.
However, conferences can be far from where many live.
Reductionist accounts can make the premise of presenting seem absurd: many will have to fly for more than 30 hours merely to make a 10 minute talk at this year's VIS---although this obviously precludes extrinsic benefits.\footnote{
  We do not mean to attack the decision to host the conference in Australia.
  There are good reasons to have the conference there: the Victorian state government has helped cover costs in a difficult time, and it is the first time the conference will be held in the southern  hemisphere (and shockingly only the 4th outside the US).
  Yet, the distance from many VIS \asLink{https://observablehq.com/d/9131e8d7a82ec721}{geo-centers} offers an extrema for reflection.
} Networking at conferences can grow the community of people familiar with your work, foster collaborations, and gain personal connections.
Academic travel is also seen as a job perk: it is fun to be paid to travel to new places around the world multiple times a year.

More quantitatively, Jacques~\cite{jacques2020chi} studied the CO$_2$ emissions from CHI, finding that the total emissions are generally increasing.
Berne \etal{}~\cite{berne2022carbon} found that flying to conferences was positively correlated with h-index.
They concluded that ``flying is a means for early-career scientists to obtain scientific visibility, and for senior scientists to maintain this visibility.''
Wynes \etal{} found that academic travel was correlated with salary (but not h-index) \cite{Wynes19Air}. This suggests that common academic metrics of success encourage us to travel to conferences.
Is visibility and an h-index bump worth significant negative climate contributions?

There are efforts by academics to fly less on a personal level \cite{Kalmus_2018,FlyingLess}, but in this paper we focus on activism at the organization level. Institutional-wide changes mitigate some of the negative impacts that would be felt on a personal level (\ie{} deciding not to fly to a conference that everyone else is going to).
Moreover, an organizational decision would affect many more people than advocating for individual choices.

\section{IEEE VIS Organizing Considerations}

Organizing the VIS conference is an immense effort undertaken by community volunteers. Many considerations go into how to choose the next location and format of the conference, many of which are not visible or widely discussed.
To better understand these deliberations, we interviewed executive-level organizers of the IEEE VIS conference about their views on conference sustainability.

Current and recent members of the VIS Executive Committee (VEC) and the VIS Steering Community (VSC) were invited via email to discuss conference logistics and to elicit their thoughts on conference sustainability.
6 organizers participated.
Participants had a total of $\approx$110 years of attending and $\approx$90 years of organizing the VIS conference.  We quote participants \qt{like so.}\footnote{To maintain anonymity of this small population, we do not release demographics or associate quotes to ids.}
Our Zoom-based interviews lasted $\approx$30 minutes and were automatically transcribed.
No formal qualitative analysis was conducted.
The set of interview questions can be found in the appendix.

While our sample size represents a third of the VEC and VSC, it is still small.
As such, we do not claim to present a comprehensive view of the concerns of all VIS organizers.
Instead, we offer a starting point for subsequent considerations and discussions.
Participants discussed a variety of concerns about the sustainability of organizing the conference, which we discuss here.

\parahead{Financial Risks}
The financial feasibility of the conference was seen as an important aspect of organizing.
One participant described it as a \qt{precondition,} and that \qt{without it, you can't do it.}
If the costs are too high or if the number of registrations is too low, the organizers run the risk of a budget deficit.
One question that arose was \qt{how to predict how many attendees will you have?} This logistical estimation is difficult, given that the number of attendees has grown over time, there is variation from year to year, and the prediction has to happen $\approx$2 years in advance.
Another consideration is that \qt{IEEE takes on that financial risk,} with VGTC also handling some of the financial aspects.

The goal is not to have the most cost-effective conference, but rather spend money where we want to support our values.
One participant mentioned the VISKids Child Care available at VIS22, saying that the organizers paid for child care to support parents.
Another participant echoed this idea, saying, \qt{if there's extra money, part of that stays at the VGTC. So the VGTC can reinvest this in... diversity scholarships... [and] accessibility.}
It was unanimously agreed that diversity and accessibility are valued at IEEE VIS, and time and money are dedicated to support these goals.
How much money are we willing to dedicate towards environmental sustainability?

\pagebreak

\parahead{Networking}
Some participants described networking as one of the essential conference goals.
This can encompass attendees' opportunities \qt{to create new ideas, also to plan about careers, next steps, invites, joint proposals.} This is reflected in the conference schedule: coffee breaks are built into the program to facilitate conversation.
One participant strongly valued such in-person interactions, saying, \qt{I don't think you get the mentoring and the education without the interpersonal interaction.}

However, not all participants saw in-person networking as essential. One questioned
\qt{Is in-person networking required for an academic career? That's probably the fundamental question.}
Prior work~\cite{berne2022carbon, heffernan2021academic} suggests that academic networks are necessary for career success, however it is unclear if physical co-location is necessary to support the formation of those networks~\cite{wassenius2023creative}.
What aspects of in-person networking are lost in virtual contexts? Are there ways to maintain some benefits of networking in other formats?

\parahead{Volunteer Labor}
VIS is primarily organized by volunteers from the community.
One participant cited the availability of volunteers for general chairs as the biggest constraint, noting that recruiting them can be especially difficult, as they only \qt{get anywhere from 1--4 bids in a year.}\footnote{One reviewer noted that the opposite is true for non-leadership positions. For example, VIS often has more applicants than it needs for student volunteer roles.}
They went on to describe location selection as boiling down to
\qt{we want to have VIS at Place X, because Person X said they will do it, and we're happy that somebody said they would do it.}

Another participant discussed volunteer labor in the context of organizational sustainability: \qt{What I find very interesting are the incentive structures. Why people engage in organizing conferences like VIS, and how we can keep up a healthy community in the sense that people volunteer to take on roles in VIS...
  That's a question also for sustainability.}
Volunteer labor may potentially be an obstacle for conference formats that require more volunteers or effort.
What are the benefits and incentives for volunteering? How can we encourage more volunteers to ensure the organizational sustainability of VIS?

\parahead{Environmental Sustainability}
Participants did not prioritize environmental sustainability over other values (such as diversity and accessibility).
Some saw it as \qt{a nice to have,} and another noted \qt{I don't know that environmentally friendly has entered into it much at all.}
However, some participants were more enthusiastic: \qt{sustainable and environmental friendly. Of course. Yeah, that's for the long term. We also need that.}

As a provocative question, we asked participants about the carbon footprint of an academic conference.
Most participants asked, \qt{Relative to what?} Relative to other computer science conferences and industry conferences, VIS is small at only $\approx$1000 people.
Other related conferences---such as \qt{NeurIPs} (13k~\cite{NeurIPS19Kurenkov}) or \qt{CHI} (3--4k~\cite{chiConferenceHistory})---have many more participants than VIS and so our impact relative to them is small.
However, the carbon cost of a round-trip international flight is immense, relative to an individual's entire footprint.
One participant suggested that \qt{academics have 3 times the carbon footprint of other people.}
Some organizers mentioned the impact of air travel. One commented, \qt{VIS has a big footprint, I would say, due to the international flights} and another commented on the location for 2023, \qt{if it is like this year in Melbourne, where it's a long haul for people in the Americas and Europe, that's a much larger footprint.}

As another provocative question, we asked our participants if VIS has a responsibility to be sustainable. Some participants agreed with this, with one saying, \qt{every conference, including VIS, has the responsibility to become more sustainable than they already are.} Understandably, some participants commented that VIS alone does not carry the responsibility, arguing that governments and other higher-level institutions (\eg{} \qt{IEEE}) should also address climate change.
One participant noted, \qt{Every conference has a responsibility to uphold reasonable standards of sustainability, right?... If you're choosing food containers... those should be compostable or reusable plates or whatever, as opposed to lots of plastic garbage.}
This participant advocated for \qt{easy choices for environmentalism,} which can include reducing food waste, not printing out conference programs, or excluding meat-based food options~\cite{ritchie2020}.
We agree that these are good choices, but we argue that now is the time for difficult choices.
Should we exchange how enjoyable the conference is or limit in-person networking to make VIS more sustainable?
Should we dedicate resources to have a less ecologically harmful conference?

\begin{figure*}[t]
  \centering
  \includegraphics[width=0.9\linewidth]{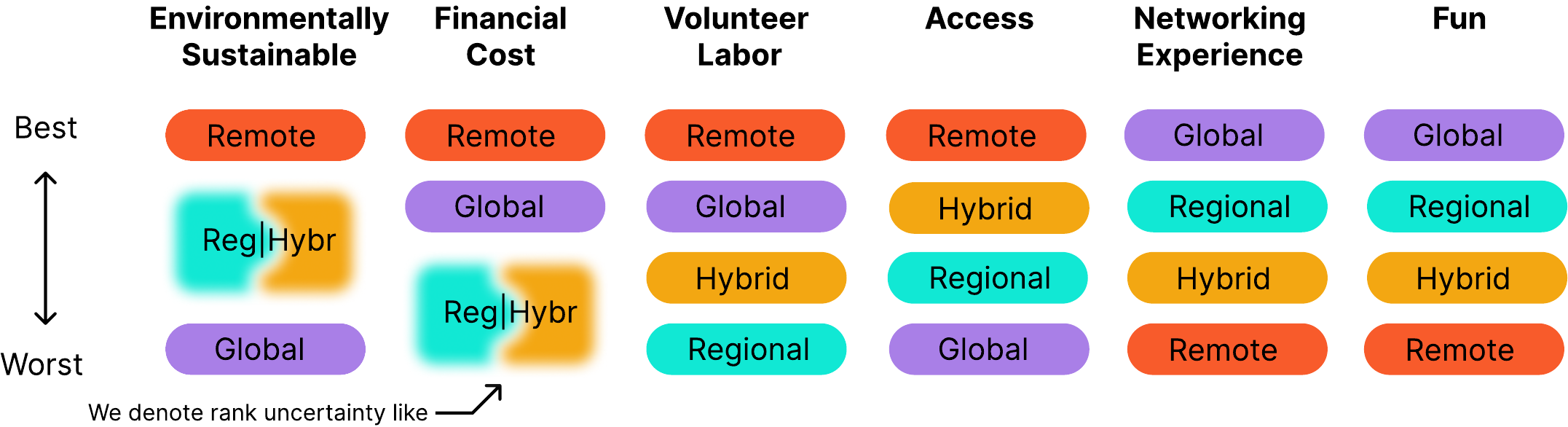}
  \vspace{-1em}
  \caption{Speculative rankings for several conference formats on various logistical considerations. Some rankings are more uncertain than others: hybrid may be more or less environmentally sustainable depending on location and virtual attendance.}
  \vspace{-1em}
  \label{fig:ranker}
\end{figure*}

\section{Discussion}
Next, we consider alternatives to the single-venue global meeting status quo.
The pandemic prompted a natural experiment in which three different formats were explored (remote at VIS20, regional satellites at VIS21, and hybrid at VIS22), each of which come with competing trade-offs.
We sketch these trade-offs in \figref{fig:ranker}.

\parahead{Remote conferences}
One of the most extreme approaches to reducing our carbon footprint would be to switch to an entirely online format.
Without taking a formal poll, we expect that this might be unpopular.
One participant went so far as to say, \qt{I'm not going to participate in the virtual conferences.}
The remote option leaves no opportunity for in-person networking, and many feel like they are less engaging than in-person conferences.
Many attendees did not feel driven (or were unable) to block off the time to become immersed in the conference in the way in-person conferences demand. One participant said, \qt{I still think we haven't found a replacement for on site or co-located networking that is satisfying.}
But, a more radical reorganization of a remote conference could be explored, such as at Info+ 2021's, wherein presenter videos were posted to the open internet prior to the conference and then conference time was used for extended discussions.

Remote conferences are substantially more accessible and equitable than in-person events.
Those who cannot afford or get visas to attend would not be excluded by their circumstances.
Attending an in-person conference can be prohibitively expensive for academics from lower-income countries and industry practitioners who do not have job-provided funding to attend.
Remote (and hybrid) formats are also friendlier to immunocompromised attendees. One participant mentioned, \qt{there's plenty of academics that have told the VEC that they're not willing to travel to conferences now, because they don't want to risk their health.}
A remote conference would also cut costs, as renting space, filling hotels, and feeding attendees would no longer be concerns.
Finally, being remote would be the most effective at reducing our carbon footprint, as travel and hosting carbon costs would be essentially eliminated.

\parahead{Hybrid conferences}
Hybrid offers compromise formats between virtual and physical. These typically (at VIS22, CHI21, and others) involve a smaller physical conference for in-person attendees and a Discord server with video streaming for remote attendees.
Hybrid options would seem to offer the best of both worlds, but this can leave those online supported only superficially while magnifying the cost and logistic burden of the conference for the organizers.

Hybrid formats were divisive.
One participant did not like the hybrid format, saying \qt{we haven't got the technology right yet to make it a good experience for everybody.}
Another participant had quite the opposite feeling about hybrid, saying that VIS22 was \qt{a huge success} and \qt{in my view,... probably one of the best conferences.}
Similarly, in personal communications, we have heard from those on the virtual end of that hybrid both that it was wonderful and that it was disastrous.
Perhaps this is an area where more software innovation is needed\footnote{Some non-academic conferences have successfully explored mediums beyond the \emph{pandemic standard} of Zoom + Discord + Gathertown. For instance, the indie-games conference Roguelike Celebration~\cite{RoguelikeCelebration} matched the ostensible theme of the venue by presenting the conference in a multi-user dungeon or MUD. MUDs are an early form of virtual world whose retro origins are similar to the game Rogue, from which Roguelike Celebration gets its name. While far from becoming a conventional approach, it shows that there is more space yet to explore.}, however it may also be necessary to rethink conference organization (such as is being explored by ICER~\cite{koICER21}).

Hybrid also necessitates substantial extra labor compared to physical or remote.
One participant highlighted the volunteer burden:
\qt{it's hard to find a general chair, and if you want to add a hybrid option. Now, I'm basically asking you as the general chair to run 2 conferences.}
Furthermore, hybrid options cost more than completely in-person or completely remote experiences.
The reduction in conference emissions depends on how many people opt to attend virtually compared to in-person.
A hybrid conference with few virtual attendees may not be more environmentally friendly than just having a single global meeting.

\parahead{Regional conferences}
An alternative compromise to a single global location is to have regional or satellite conferences (also known as multi-hub conferences \cite{parncutt2021multi}).
VIS21 orchestrated 8 informal regional events to accompany an overarching virtual edition.\footnote{However, these were quite unevenly distributed. For instance, the state of Illinois had two events (in Chicago and Urbana) while the entirety of Europe had two (in Norrk\"{o}ping and Copenhagen).}
One participant said their experience attending a satellite event \qt{was good. It was a nice event, but I think it could be more formalized.}
One approach would be to have 2--5 official locations around the world without having one central location---an idea which has also been explored by the American Association for Geographers (AAG)~\cite{aagCarbon}.
Regional conferences have proved to worked well, with FOSS4G:UK Local 2022 hosting regional events across nine venues, which was so well-received by the participants that they are planning on continuing the same format for 2023 \cite{FOSS4GUK}.
Regional events make it more feasible for people to travel semi-locally and still enjoy the in-person experience, however this limits networking to the people in your regional area.

Complicating matters is that more locations would require even more volunteers to operate. One participant mentioned this as a factor in VIS21, that \qt{they were up to volunteers to start}, while another participant saw it as a major obstacle for this format. Another participant noted that \qt{it's too much work. Somebody's got to organize that and who's gonna want to organize an event that is just a watching party?}
This view of a `watching party' may be overly specific to the VIS21 satellite model, however this is far from the only conceivable method of regionalization. Like FOSS4G:UK Local 2022, a regional conference could be a combination of streamed content and locally in-person content. Like Info+ 2021, in-person aspects of the conference could focus more on conversations and discussions.
Incentives other than tenure service requirements may be useful for prompting OC-participation (or just alternatives for those not on the tenure-track)---such as awards or money.
Some sticks may usefully match such carrots: analogous to Ko's~\cite{ko23reviewing} sustainable reviewing marketplace proposal (in which authors cannot submit papers if they have not done a corresponding amount of reviewing), service requirements could be placed on institutions to balance their usage of community resources.
There are many open questions about the logistics. For instance, how would  time zones be navigated? One participant suggested a schedule similar to the hybrid \qt{Outlier Conference,} which ran from 9am to 1am local time to accommodate other locales.
How would spreading out VIS's financial \qt{tight margins} to multiple locations work? What do \qt{legal obligations} like insurance look like for satellites?
Would organizing bodies like IEEE still support the logistics, overhead, and risk of such a reorganization?
Would this negatively silo regional communities?

\pagebreak

\parahead{An alternative cadence}
One idea that has emerged in other conferences (particularly CHI~\cite{AccessSIGCHI22}) is to limit, but not stop, the rate at which we hold international conferences.
In particular, the conference cadence could involve an international ``full'' VIS   every three years, while the other two years would be coordinated regional events (or online events \cite{bousema2020reducing}).
One participant also mentioned this idea, suggesting a global conference \qt{every N numbers of years... it becomes almost like an Olympics, like every 4 years, you have a big event, and in between you have several almost like qualifying type of events.}
In this model, presenters in the `non-global years' could be invited to share their work as a poster presentation in person at the global event.
Our participant noted that UIST has implemented something similar this year (as has PLDI~\cite{pldiProceedings}), by inviting presenters who presented their work virtually in the past three years to present again as a poster.
Such a cadence could be matched with the length of a typical Ph.D. program to allow every student the opportunity to present their work at one global conference.
We could consider a timeline of every 3 years, to account for shorter Ph.D programs (such as in the UK), compared to longer Ph.D programs (such as in the US).
This timeline could still disadvantage students without a paper ready in their first year, but could be worked around through low stakes submission formats like first year consortium's.

\parahead{(At least briefly) Consider the environmental impact of site selection}
Finding and selecting a venue for an international conference is an obviously difficult and technical process full of complex moving pieces.
Still, the environmental impact of the site could be a more central consideration in the decision-making process for VIS.
Additionally, VIS could publish the factors that go into the decision-making process for site selection, similar to CHI~\cite{ChiSiteSelection}.

However, optimizing location for emissions may negatively affect global equity.
Hosting VIS23 in Australia will likely bring local participants that would not  have joined North America-based events.
Similarly: \qt{If you look at Berlin and then look at let's say 2017 in Phoenix, you'll see a big shift between Europeans and Americans.}
One participant contrasted optimized locations with accessibility,
noting that if we \qt{make the conference central to the majority of participants} by minimizing travel distance, we impede \qt{the people with limited travel budgets, [who] lose accessibility to attend the in-person part.}
One option that we could consider is to provide travel grants for academics that live in locations disadvantaged by site selection.
Offering these travel grants might increase the amount of carbon emissions, but the goal is to view sustainability holistically among all considerations. In this way, we are increasing equity for geographical diversity, while still hosting the conference in a place that minimizes travel distance for most other attendees, and thus overall lowering the total carbon emissions.
No matter where or how we decide to host VIS, there seems to be no universal winning strategy for this \qt{multi-pronged objective function}.

\parahead{Drastic Measures}
A more drastic solution might be to decouple publishing from conferences, so that attendees are not pushed across the globe on a clockwork cadence to present their work, but rather can present  when ready at global venues (like VIS) or more local venues (like EuroVIS or PacificVIS).
This could be supported by a stronger embrace of the journal-style publishing found in every other scientific field, rather than our current conference style\footnote{While venues like TVCG and JOVI exist, subjectively, they are not the primary way in which visualization papers are published.}. Decoupling publications from conferences would allow researchers to publish papers without requiring them to also physically travel to present their work in-person.
Some recent initiatives (such as that of SIGCOMM~\cite{sigcomm}) explore this idea, however it has yet to reach the mainstream.
An even more avant-garde proposal explores removing the synchronous component of the conference entirely by mediating presentations through the mail in the form of zines~\cite{zineBasedConference}.

\parahead{Other Opportunities for Sustainability}Interest in sustainability and environmental concerns has not been seen as essential to VIS organizing. One faulty measure to this end, is that in the years since VEC/VSC became elected positions, only one candidate has made any allusion to sustainability in their bios and statements.
It may be useful to form a sustainability committee as part of the VIS OC, so as to give environmental sustainability a seat at the table.
This would be akin to the recently added Open Practices chairs.
Similar organizing bodies are present at other venues (\eg{} CHI and UIST).

Reviewers of this paper questioned why conference sustainability considerations are presented as an alt.vis paper rather than a meetup, (as one participant suggested) a VIS panel discussion, a question at the VIS town hall, or a Visualization for Social Good paper.
We believe that these forms of engagement would be welcome and likely useful.
We focused on this venue because we wanted to reach a broader swath of the community than those who might opt in to attending a more sustainability-focused event (as those attending such an event are likely to already think sustainability is valuable)---although we recognize that this venue has its own selection biases.

Several participants compared VIS to other conferences, sometimes pointing out how much bigger they are, or their offerings and formats. In interviews, VIS's small size was seen as a downside, leaving us with fewer volunteers, and the feeling that our actions are only a drop in the bucket.
We suggest that rather than making our actions inconsequential, it allows us to be nimble and able to explore new formats more easily.
Just as we compare ourselves to others, others look to us.
If we make an effort to lower our emissions, we can be an example and an inspiration for other communities.

Another means of decreasing the size of our footprint is via carbon offsets.
These offsets are ostensibly a way to purchase ``negative'' carbon emissions, such as by funding systems that capture carbon (\eg{} forests).\footnote{
  As noted above, flying from Seattle to Melbourne generates about 2.7 tons of CO$_2$~\cite{GuardianFlights}.
  To capture that much carbon, one must grow around 110 trees for a year (notably, to be effective those trees need must not have already been growing~\cite{Song2019Credits}).
  To ``offset'' the carbon emissions for the next full pre-pandemic size VIS, we must plant approximately 130k trees for a year~\cite{treesEEPA}. This is about a fifth of the trees in New York City~\cite{treesNYC}.}
There is some nuance here: carbon offsets are known to sometimes be ineffective~\cite{Astor23Credits, Song2019Credits} and increase the financial cost of the conference, but at least they would be more than nothing\footnote{The well-known mantra reduce-reuse-recycle is commonly misunderstood~\cite{Wikipedia_2023} as putting recycling on an equal foot with use reduction and the reuse of extant goods, when in fact, it is a hierarchy where reducing is more effective than reusing, which is more effective than recycling.
  We suggest that offsets have a similarly tertiary place in a different hierarchy: \textbf{Offload} (don't do things that aren't necessary), \textbf{Optimize} (make them less resource intensive as possible), and \textbf{Offset} (pay your way out of having done them).
}.
We suggest that, at a bare minimum, VIS could include offsets as part of conference registration.
There exists precedence to this effect: ACM~\cite{acmCarbon} has begun pushing its conferences to include a means for attendees to purchase offsets as part of registration.

\section{Conclusion}
Many of these observations have been made before~\cite{FootprintLevine, berne2022carbon, wassenius2023creative}, but we use this paper to bring the discussion to our community.
The best time to make VIS sustainable was 20 years ago. The second best time is now.
We recognize that a lot of us have taken a lot of flights in our lives to publish our work and advance our careers.
Rather than focus on our past decisions, we can push our community to take sustainable steps forward.
Our world is crumbling into rising oceans and forest wildfires. But there is hope: it has not crumbled yet.
Pandemic-era innovations illuminated ways for us to be more sustainable, but we are slowly slinking back to the status quo of in-person conferences.
As a community, we can choose not to forget these lessons, we can embrace environmental sustainability as a value, and work towards making VIS more sustainable.

\section{Acknowledgments}

Thank you to Eytan Adar, Cliff Lampe, Kentaro Toyama, and all of our VEC and VSC participants for sharing their thoughts on conference organizing.

\bibliographystyle{abbrv-doi}

\bibliography{bibliography}

\pagebreak
\appendix{}

\section{Interview instrument}

Here we provide the questions involved in our interview. A pdf of the specific Qualtrics form used is available at \osf{}.

\begin{enumerate}
  \item How many times have you attended the VIS conference (approximately)?
  \item How many times have you helped with organizing the VIS conference?
  \item Please rank the following items by how important they are when planning the logistical aspects of the VIS conference.
        \begin{itemize}
          \item The event should be financially responsible.
          \item The event should be accessible for people with disabilities.
          \item The event should allow for networking opportunities and support professional connections.
          \item The event should be an enjoyable and engaging conference experience.
          \item The event should be sustainable and environmentally friendly.
          \item The event should support diversity and inclusion.
        \end{itemize}
  \item What are other logistical considerations for planning VIS that are important to you? (Verbal response)
  \item Please explain why you ranked those items in that way (verbal response).
  \item The carbon footprint of an academic conference is (multiple choice)
        \begin{itemize}
          \item Immense
          \item Large
          \item Medium
          \item Small
          \item Tiny
        \end{itemize}
  \item Please explain your previous answer (verbal response).
  \item  Please rate your agreement with the following statement: VIS has a responsibility to be a sustainable conference
        \begin{itemize}
          \item Strongly disagree
          \item Disagree
          \item Slightly disagree
          \item Neither agree nor disagree
          \item Slightly agree
          \item Agree
          \item Strongly agree
        \end{itemize}
  \item Please explain your reasoning for the previous question (verbal response).
  \item Is there anything else that you wanted to mention? (verbal response)
\end{enumerate}

\end{document}